\documentclass[aps,pra,preprint,showpacs,floatfix]{revtex4}
\usepackage{graphicx}

\begin{document}

\title{Realistic theory of electromagnetically-induced transparency and slow light in a hot vapor of atoms undergoing collisions}

\author{Joyee Ghosh, R. Ghosh\cite{rg}}
\affiliation{School of Physical Sciences, Jawaharlal Nehru
University, New Delhi 110067, India}

\author{F. Goldfarb, J.-L. Le Gou\"et and F. Bretenaker}
\affiliation{Laboratoire Aim\'e Cotton, CNRS - Universit\'e Paris Sud 11, 91405 Orsay Cedex, France}

\date{20 January 2009}

\begin{abstract}
We present a realistic theoretical treatment of a three-level $\Lambda$ system in a hot atomic vapor interacting with a coupling and a probe
field of arbitrary strengths, leading to  electromagnetically-induced transparency and slow light under the two-photon resonance condition.
We take into account all the relevant decoherence processes including collisions. Velocity-changing collisions (VCCs) are modeled in the
strong collision limit effectively, which helps in achieving optical pumping by the coupling beam across the entire Doppler profile. The
steady-state expressions for the atomic density-matrix elements are numerically evaluated to yield the experimentally measured response
characteristics. The predictions, taking into account a dynamic rate of influx of atoms in the two lower levels of the $\Lambda$, are in
excellent agreement with the reported experimental results for $^4$He*. The role played by the VCC parameter is seen to be distinct from
that by the transit time or Raman coherence decay rate.
\end{abstract}


\maketitle

\section{Introduction}
Electromagnetically-induced transparency (EIT) in three-level $\Lambda$-systems is a phenomenon in which an initially absorbing medium is
rendered transparent to a resonant weak probe laser when a strong coupling laser is applied to a second transition \cite{harris91}. It is
based on quantum interference effects involving coherence between the two lower states of the $\Lambda$. The quest has been on for simple
room-temperature systems capable of demonstrating EIT for applications ranging from narrow transparencies to switchable and controlled
broadband delays, and slowing of light \cite{kash99} for use in quantum-information processing. Room-temperature atomic vapors have been
found to be attractive candidates for such applications, necessitating a treatment of the phenomenon taking into account effects of atomic
motion and collisions.

There are some early studies \cite{arimondo_96} of Doppler-broadening effects in EIT for a system of moving atoms. In the limit of vanishing
probe field and under the assumption that all atoms are trapped in the dark state (which is a coherent superposition of the two lower levels
of the $\Lambda$), it was found that power-broadening of the EIT linewidth takes place: $\Gamma_{\mathrm{EIT}} = \Omega_{\mathrm{C}}^2/4
W_{\mathrm{D}}$ (where $\Omega_{\mathrm{C}}$ is the Rabi frequency of the coupling field and $W_{\mathrm{D}}$ is the Doppler half-width at
half-maximum), which is similar to the well-known result for a homogeneously-broadened system: $\Gamma_{\mathrm{EIT}} =
\Omega_{\mathrm{C}}^2/4 \Gamma_0$ (where $\Gamma_0$ is the homogeneous linewidth). This dependence was experimentally verified in
Ref.\,\cite{kash99}. In the limit of relatively low probe field intensity, $\Omega_{\mathrm{P}} \ll (\Gamma_0
/W_{\mathrm{D}})\Omega_{\mathrm{C}}$, and under the same assumption of full coherent trapping (i.e., neglecting the two-photon coherence
decay), other workers \cite{yudin00} have derived the following result for the EIT linewidth: $\Gamma_{\mathrm{EIT}} = \Omega_{\mathrm{P}}
\Omega_{\mathrm{C}}/4 \Gamma_0$, where $\Omega_{\mathrm{P}}$ is the Rabi frequency of the probe field.

For EIT in a room-temperature gas, the decay of two-photon (Raman) coherence is caused by several mechanisms, such as transit-time
broadening, population exchange, and atom-atom and atom-wall collisions. Insight into the most significant decoherence mechanism can be
gained by measuring the width of the EIT resonance as a function of the coupling field intensity. An existing theoretical treatment of EIT
in Doppler-broadened gases \cite{javan02}, assuming the population exchange between the lower levels to be the main source of decoherence,
predicts a {\it nonlinear} dependence of the EIT width $\Gamma_{\mathrm{EIT}}$ on the coupling beam intensity for weak coupling powers. In
the limit of very large coupling intensity, it is shown to reduce to the power-broadening case. Javan {\it et al}. \cite{javan02} consider a
closed atomic model scheme and argue that such a model gives a description almost equivalent to the one for an open system in which atoms
decay (out of the interaction region) with the rate $\Gamma_{\mathrm{R}}$, and atoms come into the interaction region with equally populated
lower levels. Though it is a theory for EIT in a Doppler-broadened medium, the role of collisions is completely neglected in
Ref.\,\cite{javan02}.

Most existing experiments in atomic vapors \cite{expt_vapor, lvovsky06, goldfarb08} have shown the dependence of the width of the EIT
resonance on the coupling field intensity ($\propto \Omega_{\mathrm{C}}^2$) to be {\it linear}, even for weak coupling powers (with the
exception of the work by Ye and Zibrov \cite{ye02} which was performed without a buffer gas and with a very small beam diameter). As an
example, we focus on the data of our recent demonstration \cite{goldfarb08} that metastable $^4$He (He*) at room temperature is a simple
system capable of yielding EIT and slow light. For the coupling field strengths used, $\Gamma_{\mathrm{EIT}}$ is expected to evolve linearly
with $\Omega_{\mathrm{C}}$ according to Ref.\,\cite{javan02}, with a slope depending on $\Gamma_{\mathrm{R}}$. The lower-level relaxation in
a gas is mainly determined by the transit time of the atoms through the laser beam -- different beam sizes would lead to different transit
times and hence different values of $\Gamma_{\mathrm{R}}$. Our experimental results \cite{goldfarb08} have clearly shown that (i)
$\Gamma_{\mathrm{EIT}}$ evolves quadratically with $\Omega_{\mathrm{C}}$, and (ii) the slope of this evolution is the same for different
beam sizes, i.e., for different values of $\Gamma_{\mathrm{R}}$. The suggestive role played by collisions between metastable and
ground-state atoms in yielding this experimental result is not described by the theory of Ref.\,\cite{javan02}.

If all the atoms across the entire Doppler profile are assumed to be initially optically pumped by the coupling beam to the probe ground
level, a calculation \cite{lvovsky06} of the response of the medium up to first order in the probe field leads to a linear dependence of the
EIT linewidth on the coupling beam intensity \cite{approx}. To obtain this result, one also supposes that the decoherence in the lower
states is caused by pure dephasing, contrary to the assumptions of population exchange in Ref.\,\cite{javan02}. In a realistic situation,
the special initial condition of the entire atomic population being in the probe ground level does not hold good -- the population is
equally likely to be in the probe and the coupling ground levels initially. Thus, population exchange between the two lower levels cannot be
ignored. Also, a treatment in first (linear) order in the probe field cannot possibly give results when the coupling field is small, viz. of
the order of the probe field.

In this paper, we address this deficiency in the existing theory of EIT and slow light in a hot atomic vapor, and attempt a complete,
realistic analysis taking into account all the relevant decoherence processes, for arbitrary strengths of the probe and the coupling fields,
and without assumptions of any special initial condition. We consider the influx of fresh atoms in the lower levels and the outflux from all
the levels at a diffusive transit rate in the gas. We allow for unequal rates of influx in the lower levels to take into account optical
pumping by the control field and return of coherently-prepared atoms into the interaction region. The phase-interrupting and
velocity-changing collisions (VCCs) of active atoms are also modeled effectively.

Apart from the transparency width discussed so far, there are other features of interest associated with EIT. For non-zero detunings of the
coupling field from the center of the Doppler-broadened transition frequency, the transmitted intensity profiles become asymmetric about the
two-photon resonance (Raman detuning = 0) \cite{lounis92,wong04}. This Fano-like feature is a signature of the two-photon process of EIT,
and emerges naturally from our model. The narrow spectral hole in the absorption profile of the EIT medium is accompanied by a strong
dispersion of the index of refraction according to the Kramers-Kronig relations, inducing a low group velocity. The evolutions of the peak
transmission and the group delay with the coupling beam intensity predicted from our analysis faithfully reproduce the experimentally
observed behaviors.

The paper is organized as follows. In Sec.\,II, all the different relaxation processes for EIT in a three-level $\Lambda$-system are
discussed. The fraction of atoms that come back to the beam from outside being still coherently-prepared are suitably modeled in this
section. In Sec.\,III, VCCs are dealt with separately, and the density matrix equations are written with various relaxations including that
due to VCCs. The steady-state solutions for the density matrix elements are presented in the strong-collision approximation with a model for
the VCCs under rapid VCC coverage. This is followed by our theoretical results in Sec.\,IV on the Doppler-averaged Fano-like EIT profiles,
the variation of the EIT width, the peak transmission and the group delay with the coupling intensity, all of which agree very well with the
experimental data for the He* system. The general dependence of these features on the VCC parameter, the unequal atomic influx parameter,
the Raman decoherence rate and the initial transmission are also probed. It is shown that the unequal feeding back of atoms into the lower
levels, more being in the coherently-prepared dark state $\vert b \rangle$, has an important effect on various characteristics of an EIT
medium. It is also shown that the impact of VCCs is distinct and cannot be incorporated by simply modifying the transit time decay (and
hence the Raman coherence relaxation) rate. The conclusions are presented in Sec.\,V.

\section{EIT scheme with relaxation processes}

\subsection{Level scheme for EIT}

\begin{figure}[htbp]
\centering
\includegraphics[width=0.4\textwidth]{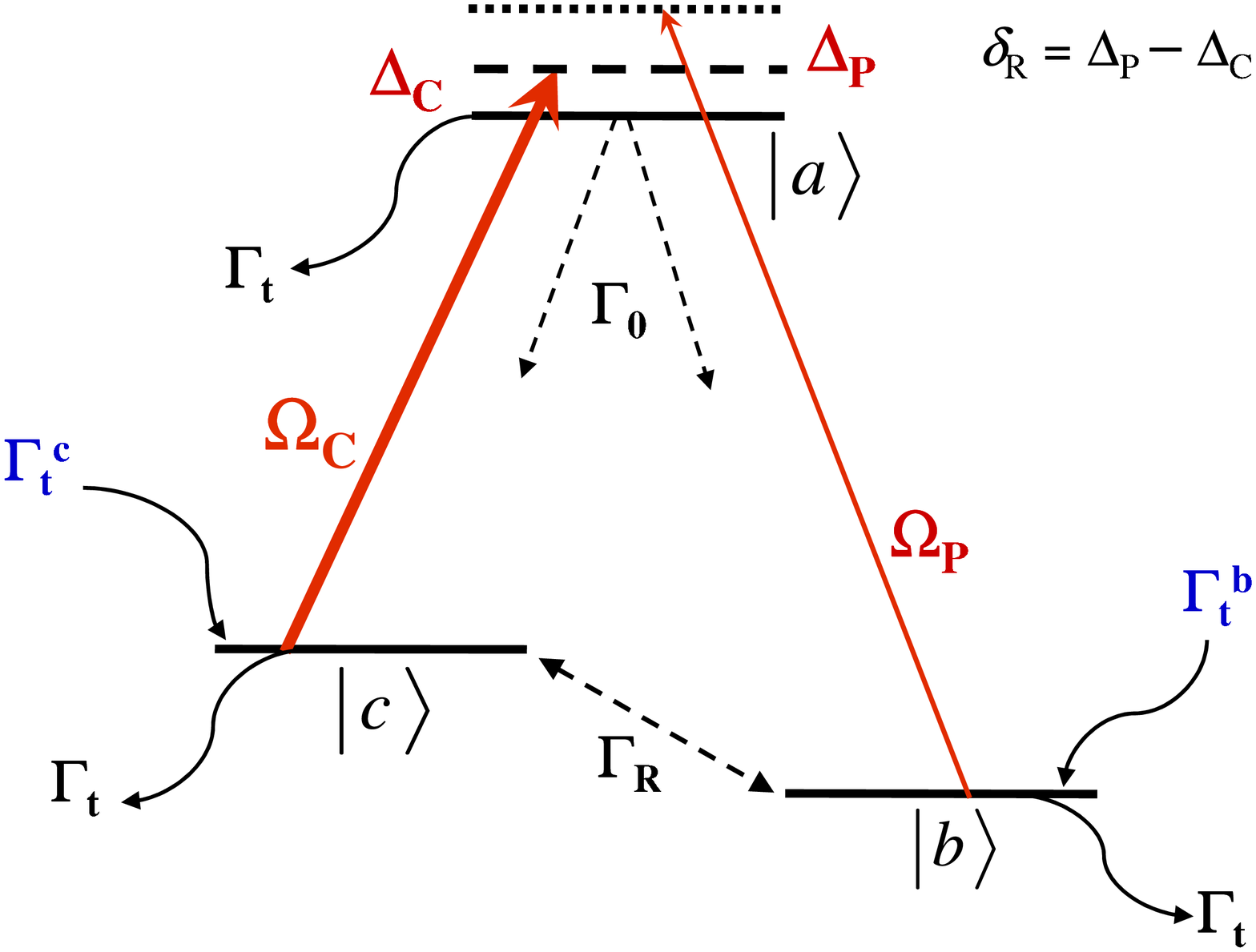}
\caption{Three-level $\Lambda$ scheme for EIT}
\label{lambda}
\end{figure}

Consider a $\Lambda$ atomic system as in Fig.\,\ref{lambda}. Levels $\vert a \rangle$ and $\vert b \rangle$ are coupled by a weak probe
field, the interaction energy being given by its Rabi frequency $\Omega_{\mathrm{P}}$. Another strong coupling field of Rabi frequency
$\Omega_{\mathrm{C}}$ couples the same excited level $\vert a \rangle$ with level $\vert c \rangle$ along the other arm of the $\Lambda$.
Both fields are treated classically. The probe detuning is
\[ \Delta_{\mathrm{P}} = \omega_{\mathrm{P}} - \omega_{ab} , \]
with $\omega_{\mathrm{P}}$ as the probe frequency, and likewise, the coupling field detuning is
\[ \Delta_{\mathrm{C}} = \omega_{\mathrm{C}} - \omega_{ac} , \]
with $\omega_{\mathrm{C}}$ as the coupling frequency. The Raman detuning is
\[ \delta_{\mathrm{R}} = \Delta_{\mathrm{P}} - \Delta_{\mathrm{C}} = \omega_{\mathrm{P}} - \omega_{\mathrm{C}} -(\omega_{ab}-\omega_{ac}). \]
The probe and the coupling fields propagate in the same direction, and the frequency difference $|\omega_{ab} - \omega_{ac}|$ is small
enough so that the residual Doppler shift $|k_{\mathrm{P}} - k_{\mathrm{C}}| v$ can be ignored \cite{residual}. For EIT, the system should
be prepared by optical pumping so that the initial population is entirely concentrated in the dark state $|b \rangle$. The coupling field
creates a quantum interference between the probability amplitudes of transition $\vert b \rangle$ $\rightarrow$ $\vert a \rangle$ via two
different channels, (i) a direct absorption process from $\vert b \rangle$ to $\vert a \rangle$, and (ii) an indirect stimulated Raman
process from $\vert b \rangle$ to $\vert a \rangle$ to $\vert c \rangle$ to $\vert a \rangle$. Under the appropriate condition of two-photon
resonance $\delta_{\mathrm{R}}$ = 0, the medium becomes effectively transparent (zero absorption) for the probe field, leading to EIT.

In the example of $^4$He, the first excited state 2$^3$S$_1$ is a metastable state with a lifetime of approximately 8000 s. The transition
between 2$^3$S$_1$ and the second excited state 2$^3$P at a gap of 277 THz or 1083 nm is conveniently used. The levels 2$^3$S$_1$ and
2$^3$P$_1$ are each split into three sublevels with $m_J$ = $-$1, 0 and 1. Out of these six levels, a $\Lambda$ system is carved out by
optical pumping (with the help of proper choice of polarization of light, selection rule and allowed stimulated/spontaneous transitions)
\cite{gilles01}, with $m_J$ = $-$1 and 1 of 2$^3$S$_1$ forming the two lower levels, and $m_J$ = 0 of 2$^3$P$_1$ the upper level. At zero
magnetic field the lower levels are degenerate and $\omega_{ab}=\omega_{ac}$.

\subsection{Relaxation processes}

The dispersion and absorption of the medium with respect to the probe field of Rabi frequency $\Omega_{\mathrm{P}}$ are determined by the
off-diagonal element $\rho_{ab}$ of the atomic density matrix. This describes the atomic coherence or the atomic polarization. Hence, it is
important to investigate the different relaxation processes affecting the optical coherence as well as the Raman coherence of the two lower
levels that lead to the dark state and EIT. It is known that collisions of active atoms, which only perturb the phase or amplitude of an
oscillating atom without changing its velocity, lead to homogeneous line-broadening and a shift of its line center. But collisions can also
result in changes in the velocity of active atoms, in addition to being phase-interrupting, and affect atomic coherences of the system under
consideration \cite{berman84}.

The present theoretical analysis examines the density matrix equations for a three-level $\Lambda$-system interacting with two fields, in
the presence of the following different sources of relaxation:
\begin{enumerate}
\item{the spontaneous decay from the excited state $\vert a \rangle$ transfers atoms to the ground states with equal decay rates $\Gamma_0/2$,}
\item{the transit of the atoms through the laser beam at a rate $\Gamma_{\mathrm{t}}$ replaces atoms (in all states) in the laser interaction
region by fresh atoms arriving (only in the lower states) from the volume outside that region. The transit rate depends on the pressure
(diffusion coefficient) of the gas. All the populations and coherences are affected because of this motion. But it is the lower-state
relaxation that is mainly determined by the transit rate $\Gamma_{\mathrm{t}}$,}
\item{the collisions (phase-interrupting) damp the atomic polarizations and coherences,}
\item{the collisions also change velocities of the active atoms without changing their internal state.}
\end{enumerate}

In our example of the He* system, the atom-wall collisions are not very significant for the dynamics. De-excitation of He* on collision with
cell walls leaves inert He atoms in the ground state (1$^1$S$_0$), which cannot be detected: there are no background atoms to contribute to
noise, unlike experiments which use true ground-state atoms such as the alkali metals. Thus, with our system, there is an advantage of
collisions of the active atoms with the walls of the cell which result in quenching.

The decay for the optical coherences is:
\begin{equation}
\frac{\Gamma}{2} = \frac{\Gamma_0}{2} + \frac{\Gamma_{\mathrm{coll}}}{2} + \Gamma_{\mathrm{t}}, \label{optcoh}
\end{equation}
where $\Gamma_{\mathrm{coll}}$ is the collisional dephasing due to pressure, and the Raman coherence decay for $\tilde{\rho}_{cb}(v)$ is:
\begin{equation}
\Gamma_{\mathrm{R}} = \Gamma_{\mathrm{t}} + {\Gamma^p}_{\mathrm{coh}} + \Gamma_{\mathrm{B}}, \label{Ramancoh}
\end{equation}
where ${\Gamma^p}_{\mathrm{coh}}$ is the collisional term proportional to gas pressure, and $\Gamma_{\mathrm{B}}$ is the dephasing due to
inhomogeneity in residual magnetic field and other possible dephasing mechanisms.

The mean free-path of the He* atoms is, in a hard sphere model, of the order of 0.1 mm. If we consider that the atoms cross the beam in a
one-dimensional random walk, we can see that, at 300 K, they experience about 10$^4$ collisions during their trip across a 1-cm-diameter
beam, leading to a diffusive transit time $\Gamma_{\mathrm{t}}^{-1}$ of the order of 0.5 ms. According to a rigorous calculation by
Fitzsimmons \cite{fitzsimmons68}, the diffusion constant at 1 Torr and 300 K is $D$ = 500 cm$^2$/s. Thus, in a 1-D diffusion model, the
variance in the displacement is $(\Delta x)^2 = 2 D t$. For $\Delta x$ = 1 cm, we again obtain the transit time, $t$ = 1 ms. For metastable
helium at room temperature and pressure of 1 Torr inside a cylinder shielded with $\mu$-metal to avoid stray magnetic fields, the various
decay rates are typically:
\begin{eqnarray*}
\Gamma_0 &=& 10^7~ \mbox{s}^{-1}, \nonumber \\
\Gamma_{\mathrm{coll}} &=& 1.33 \times 10^8~ \mbox{s}^{-1}, \nonumber \\
\Gamma_{\mathrm{t}} & \sim & 10^3~ \mbox{s}^{-1}, \nonumber \\
\Gamma_{\mathrm{R}} &=& 10^4 - 10^5~ \mbox{s}^{-1}.
\end{eqnarray*}

The coherences between the lower levels may benefit from the fact that the atoms can diffuse out of the interaction region and return before
decohering. The rates at which the atoms return to the lower states $\vert b \rangle$ and $\vert c \rangle$ are not likely to be equal,
since inside the beam, the populations in these two states are made unequal by the control-field optical pumping from $|c \rangle$ to $|b
\rangle$, and hence the population diffusing outside is also likely to be unbalanced. The rate of return from outside cannot be a constant
but would depend dynamically on the population difference, $(\rho_{bb}-\rho_{cc})$. This feature is incorporated by using unequal influx
rates:
\begin{eqnarray*}
\left.\frac{\mathrm{d}\rho_{bb}}{\mathrm{dt}}\right|_{\mathrm{in}} &=& \frac{\Gamma_{\mathrm{t}}}{2} \left[ 1 + \beta \left( \rho_{bb} - \rho_{cc} \right) \right], \\
\left.\frac{\mathrm{d}\rho_{cc}}{\mathrm{dt}}\right|_{\mathrm{in}} &=& \frac{\Gamma_{\mathrm{t}}}{2} \left[ 1 - \beta \left( \rho_{bb} - \rho_{cc} \right)\right],
\end{eqnarray*}
while maintaining a single departure rate of $\Gamma_{\mathrm{t}}$ from the beam in all the states. A value of $\beta = 1$ would indicate
that all atoms going back to the beam are prepared for EIT in the dark state $\vert b \rangle$. On the other hand, $\beta$ = 0 would
indicate that equal number of atoms enter the beam in states $\vert b \rangle$ and $\vert c \rangle$. The overall atomic population is, of
course, conserved. A physical picture of $\beta$ can be thought as arising from the treatment of the atomic motion outside the laser beam in
a diffusion equation, by assuming a random distribution of the durations spent by the atoms outside the interaction region \cite{Ramsey}.

\section{Atomic Density Matrix Formulation}

\subsection{Velocity-changing collisions}

Velocity-changing collisions (VCCs) which shuffle atoms between different velocity classes represent an important source of relaxation for
the lower states. It can modify the atomic velocity without affecting the atomic coherence in the lower states: in this case the atoms
prepared by the laser radiation in a dark state are transferred to other velocity classes \cite{arimondo96}. If this transfer applies with
high efficiency, all the atomic velocity classes are pumped into the dark state, either by direct pumping or by VCC \cite{goldfarb08}.

Our active atoms are in a three-level $\Lambda$ configuration. It is taken that collisions do not possess sufficient energy to induce
transitions between the upper and the lower levels. This assumption effectively allows one to treat the scattering of each active atom
separately, such that one can apply standard quantum mechanical scattering theory using a different total energy for each active atom level.

We use the \textit{impact approximation} in which the active atom-perturber atom collisions are viewed to occur instantaneously, i.e., the
duration $\tau_{\mathrm{c}}$ of the typical collision is assumed to be much less than the various time scales in the problem (with the
exception of the optical period $2 \pi /\omega$, where $\omega$ is a transition frequency). In particular, the assumption that
$\tau_{\mathrm{c}}$ is much lesser than the time $T_{\mathrm{c}}$ between collisions is called the \textit{binary collision approximation}.
In our example of metastable $^4$He, the buffer or perturber atoms are the $^4$He atoms in the ground (1$^1$S$_0$) state. Typically,
$T_{\mathrm{c}} \approx 10^{-7}$ s at 1.0 Torr of buffer gas pressure and $\tau_{\mathrm{c}} \approx 10^{-12}$ s. The binary collision
approximation is valid easily up to a gas pressure of 500 Torr. The implications of the impact approximation are easily understood. Each
collision produces a change in the density matrix $\rho$ associated with active atoms. Between collisions, the external field produces a
time rate of change for the density matrix. Since changes produced by the external fields during collision are assumed to be negligible, the
impact approximation allows one to represent the contributions to $d\rho/dt$ as arising independently from the collisions and external
radiation fields. In this way, one obtains a `master' or transport equation for $\rho$ \cite{rg93}.

The influence of VCC on EIT is formally treated by writing the density matrix equations for each velocity class and introducing in those
equations the probability for VCC between different classes \cite{rg93, graf95}. The contribution of VCCs to the density-matrix equations
for the populations and coherences, in general, is
\begin{equation}
\frac{d}{dt} [\rho_{ij}(v)]_{\mathrm{VCC}} = - \Gamma_{ij\mathrm{VCC}} \rho_{ij}(v) + \int K_{ij}(v' \rightarrow v) \rho_{ij}(v') dv'. \label{VCCgen}
\end{equation}
Here the VCC process is described in terms of the collisional relaxation rates $\Gamma_{ij\mathrm{VCC}}$, and the associated collisional
kernels are $K_{ij}(v' \rightarrow v)$. The collision kernel $K_{ij}(v' \rightarrow v)$ [of the dimension of inverse length] gives the
probability density per unit time that a collision changes the velocity of an active atom in state $i$ from $v'$ to $v$. Changes in $v$
occur at some average rate $\Gamma_{ij\mathrm{VCC}}$, which is related to the kernel in the following way \cite{GammaVCC}:
\begin{equation}
\Gamma_{ij\mathrm{VCC}}(v) \geq \int dv'~ K_{ij}(v \rightarrow v'). \label{eq7}
\end{equation}
The first term on the right-hand side in Eq.\,(\ref{VCCgen}) can be viewed as the ``out term" resulting from collisions that remove active
atoms in state $i$ from the velocity subclass $v$, and the last term is the ``in term" bringing atoms from other velocity subclasses into
the subclass $v$. $K_{ii}$ is related to the differential scattering cross-section, and $\Gamma_{ii\mathrm{VCC}}$ is related to the
corresponding total scattering cross-section.

Since the collision interaction depends on the internal atomic states of the atoms, a complete description requires a separate collision
kernel for each atomic state population and each coherence. In general, an excited atom has a slightly larger collisional cross-section than
a ground-state atom because excited atoms are bigger than the ground-state atoms, and therefore the excited atoms suffer a stronger
collisional damping of speeds. The opposite holds true for the case of active alkali atoms colliding with neon where an excited atom has a
slightly lower collisional cross-section than a ground-state atom. However, in a low-pressure regime where the decay rate $\Gamma$
($\approx$ 10$^8$ s$^{-1}$) at the optical transition ($|a \rangle \rightarrow |b \rangle$ or $|a \rangle \rightarrow |c \rangle$) is larger
than the collision rate $\Gamma_{0,1~ \mathrm{VCC}}$ ($\approx$ 10$^7 ~\mathrm{s}^{-1}$), the transport of coherence in the optical
transition from one velocity group to another is of not much importance. For the lower-state coherence, the situation is different, as the
effective lower state relaxation ($\approx$ 10$^4$ s$^{-1}$) is smaller than the collision rate. We can thus assume that the collision
kernel is zero for the optical coherences, is approximately the same for all populations, and is different for the lower-state coherence:
\begin{eqnarray}
K_{ab} (v \rightarrow v') &=& K_{ac} (v \rightarrow v') = 0 , \nonumber \\ K_{aa}(v \rightarrow v') &=& K_{bb} (v \rightarrow v') = K_{cc} (v \rightarrow v') \equiv K_0 (v \rightarrow v') , \nonumber \\
K_{bc} (v \rightarrow v') &\equiv& K_1 (v \rightarrow v') .
\end{eqnarray}
Thus, for the three-level $\Lambda$ system, the added VCC contributions are:
\begin{eqnarray}
\frac{d}{dt} [\rho_{ii}(v)]_{\mathrm{VCC}} &=& - \Gamma_{0\mathrm{VCC}} \rho_{ii}(v) + \int K_0(v' \rightarrow v) \rho_{ii}(v')~ dv', ~~(i = a, b, c) , \label{eq8} \\
\frac{d}{dt} [\tilde{\rho}_{cb}(v)]_{\mathrm{VCC}} &=& - \Gamma_{1\mathrm{VCC}} \tilde{\rho}_{cb}(v) + \int K_1(v' \rightarrow v) \tilde{\rho}_{cb}(v') dv' , \label{eq9}
\end{eqnarray}
where $\Gamma_{ii\mathrm{VCC}} \equiv \Gamma_{0\mathrm{VCC}}$, $\Gamma_{bc\mathrm{VCC}}=\Gamma_{cb\mathrm{VCC}} \equiv \Gamma_{1\mathrm{VCC}}$, and $\Gamma_{ab\mathrm{VCC}}=\Gamma_{ba\mathrm{VCC}}=0$. Here we have further neglected the velocity-dependence of the collision rate $\Gamma_{0,1~ \mathrm{VCC}}$ because it is usually a slowly-varying function of velocity.

\subsection{Complete set of density matrix equations with relaxations}

We first convert the usual density matrix elements $\rho_{ij}$ to slowly-varying variables $\tilde{\rho}_{ij}$ in order to remove the fast
optical oscillations by using the following transformations:
\begin{eqnarray}
\rho_{ab} &=& \tilde{\rho}_{ab} e^{-i \omega_{\mathrm{P}} t}, \\
\rho_{ac} &=& \tilde{\rho}_{ac} e^{-i \omega_{\mathrm{C}} t}, \\
\rho_{cb} &=& \tilde{\rho}_{cb} e^{-i (\omega_{\mathrm{P}} - \omega_{\mathrm{C}}) t}.
\end{eqnarray}
Taking into account the various relaxations mentioned above, including the dynamic atomic influx into the beam, and incorporating the effect
of VCCs expressed in Eqs.\,(\ref{eq8})-(\ref{eq9}), the evolution of the slowly-varying density matrix elements $\tilde{\rho}_{ij}(v)$ for
atoms with velocity $v$ may be written under the rotating-wave approximation as:

\begin{eqnarray}
\frac{d \rho_{aa}(v)}{dt} &=& - (\Gamma_0 + \Gamma_{\mathrm{t}} + \Gamma_{0\mathrm{VCC}}) \rho_{aa}(v) - i\frac{\Omega_{\mathrm{P}}}{2} [\tilde{\rho}_{ab}(v) - \tilde{\rho}_{ba}(v)] \nonumber \\ && - i\frac{\Omega_{\mathrm{C}}}{2} [\tilde{\rho}_{ac}(v) - \tilde{\rho}_{ca}(v)] + \int K_0(v' \rightarrow v) \rho_{aa}(v')~ dv', \label{eq14} \\
\frac{d \rho_{bb}(v)}{dt} &=& \frac{\Gamma_0}{2} \rho_{aa}(v) - \left( \Gamma_{\mathrm{t}} + \Gamma_{0\mathrm{VCC}} \right) \rho_{bb}(v) + \frac{\Gamma_{\mathrm{t}}}{2} \left[ W(v) + \beta \left( \rho_{bb}(v) - \rho_{cc}(v) \right) \right] \\
&& + i\frac{\Omega_{\mathrm{P}}}{2} [ \tilde{\rho}_{ab}(v) - \tilde{\rho}_{ba}(v)] \nonumber + \int K_0(v' \rightarrow v) \rho_{bb}(v')~ dv', \label{eq15} \\
\frac{d \rho_{cc}(v)}{dt} &=& \frac{\Gamma_0}{2} \rho_{aa}(v) - \left( \Gamma_{\mathrm{t}} + \Gamma_{0\mathrm{VCC}} \right) \rho_{cc}(v) + \frac{\Gamma_{\mathrm{t}}}{2} \left[ W(v) - \beta \left( \rho_{bb}(v) - \rho_{cc}(v) \right) \right] \\ && + i\frac{\Omega_{\mathrm{C}}}{2} [ \tilde{\rho}_{ac}(v) - \tilde{\rho}_{ca}(v)] \nonumber + \int K_0(v' \rightarrow v) \rho_{cc}(v')~ dv', \label{eq16} \\
\frac{d \tilde{\rho}_{ab}(v)}{dt} &=& - \left[ \frac{\Gamma}{2} - i (\Delta_{\mathrm{P}} - k v) \right] \tilde{\rho}_{ab}(v) + i\frac{\Omega_{\mathrm{C}}}{2} \tilde{\rho}_{cb}(v) - i\frac{\Omega_{\mathrm{P}}}{2} [\rho_{aa}(v) - \rho_{bb}(v)], \label{eq17} \\
\frac{d \tilde{\rho}_{ca}(v)}{dt} &=& - \left[ \frac{\Gamma}{2} + i (\Delta_{\mathrm{C}} - k v) \right] \tilde{\rho}_{ca}(v) - i\frac{\Omega_{\mathrm{P}}}{2} \tilde{\rho}_{cb}(v) + i\frac{\Omega_{\mathrm{C}}}{2} [\rho_{aa}(v) - \rho_{cc}(v)], \label{eq18} \\
\frac{d \tilde{\rho}_{cb}(v)}{dt} &=& - \left[ \Gamma_{\mathrm{R}} + \Gamma_{1\mathrm{VCC}} - i \delta_{\mathrm{R}} \right] \tilde{\rho}_{cb}(v) - i\frac{\Omega_{\mathrm{P}}}{2} \tilde{\rho}_{ca}(v) + i\frac{\Omega_{\mathrm{C}}}{2} \tilde{\rho}_{ab}(v) \nonumber \\ && + \int K_1(v' \rightarrow v) \tilde{\rho}_{cb}(v')~ dv'. \label{eq19}
\end{eqnarray}
Note that the {\it total} atomic population in state $i$ at a time $t$ is given by
\begin{equation}
\rho_{ii} = \int_{-\infty}^{\infty} \rho_{ii} (v) ~dv,
\end{equation}
and $\sum_{i=1}^3 \rho_{ii} (t) = 1$ for a closed system. $\tilde{\rho}_{ij}(v)$s have the dimensions of inverse speed. Equations
(\ref{eq14})-(\ref{eq19}) are to be solved for $\tilde{\rho}_{ab}$ to get the susceptibility at $\omega_{\mathrm{P}}$.

\subsection{Strong collision approximation}

These integro-differential equations can be solved using iterative techniques which may be taken up in future. The solution can be worked
out for various limiting forms of the collision kernel, and here we follow the strong collision model, in which $v(t)$ is a jump process.
The effect of collisions is `strong', i.e., it washes out the memory of the pre-collision value of the velocity. A single collision, on
average, thermalizes the velocity distribution. The rate of collisions is taken as an average rate given by the inverse of the mean
free-time between collisions. The collision kernel is then greatly simplified and can be approximated by
\begin{equation}
K_{0,1}(v' \rightarrow v) = \Gamma_{0,1~ \mathrm{VCC}}~ W(v), \label{eq11}
\end{equation}
where $W(v)$ is the Maxwell-Boltzmann distribution for the velocity vector in one direction, given by
\begin{equation}
W(v) = \frac{1}{\sqrt{\pi} u} e^{-(v/u)^2}, \label{eq12}
\end{equation}
where $u$ is the most probable speed:
\begin{equation}
u = \sqrt{\frac{2 k_{\mathrm{B}} T}{m}} , \label{eq13}
\end{equation}
$m$ being the mass of an atom. For a temperature of 300 K, the most probable speed of He atoms is about 1100 ms$^{-1}$. In the presence of
light of wave-number $k$, the 1/$e$ Doppler half-width is $ku$. For our system, with a laser at a frequency $\omega_{\mathrm{P,C}}$ = 1.74
$\times$ 10$^{15}$ rad/s (or wavelength $\lambda_{\mathrm{{P,C}}}$ = 1.083 $\mu$m), the Doppler half-width at half maximum (HWHM),
$W_{\mathrm{D}}/2 \pi$ = 0.9 GHz.

Note that under the assumption that the LHS of Eq.\,(\ref{eq11}) is independent of the initial velocity $v'$, the RHS is the only allowed
form, consistent with the detailed balance of transitions
\[ W(v') K_{0,1}(v \rightarrow v') = W(v) K_{0,1}(v' \rightarrow v) , \]
and the conservation of probability (\ref{eq7}).

\subsection{Steady-state solutions}

\subsubsection{Conditions for rapid VCC coverage}

The excitation by a single-mode laser is velocity-selective. In the absence of VCCs, optical pumping with a single-mode laser polarizes only
a small portion of the thermal velocity distribution.

In strong thermalizing VCCs, the root-mean-square velocity change $\Delta v$ is much larger than the width of the resonant velocity ``bin"
($\Delta v \gg \Gamma/k$, with $\Gamma$ the homogeneous linewidth). To ensure rapid thermalization to access the entire velocity profile,
the number of VCCs occurring during the lower-state orientation relaxation time $1/\Gamma_{\mathrm{R}}$ must be large compared to the total
number of velocity bins ($\Gamma_{0,1~ \mathrm{VCC}} /\Gamma_{\mathrm{R}} \gg 2 ku/\Gamma$) \cite{quivers86}.

Also, when the photon absorption rate $\Gamma'(v)$ is large compared to the rate of diffusion of atoms across the laser beam ($\Gamma' (v)
\gg \Gamma_{\mathrm{t}}$) but small compared to the rate of VCCs ($\Gamma'(v) \ll \Gamma_{0,1~ \mathrm{VCC}} $), the redistribution of atoms
in the different velocity classes rebuild the Maxwell-Boltzmann velocity distribution (i.e., VCCs thermalize the lower-state velocity
distributions rapidly compared to an absorption-emission cycle). This leads to a velocity-independent type of optical pumping
\cite{pritchard81}. Since atoms jump from one velocity class to another, when the number of VCCs occurring per cycle is much greater than
the number of pump photons, the pumping spreads over the entire Doppler distribution in each optical pumping cycle. The laser is depleting a
single velocity bin at a rate $\Gamma'(v)$, but since the rate $\Gamma_{0,1~ \mathrm{VCC}}$ at which the VCCs are replenishing it with atoms
from the entire Doppler distribution is so much faster, the velocity distribution stays thermalized even during the pumping process.

Under the above conditions and considering the broadening due to the thermal velocity distribution of the atoms, we re-write the density
matrix equations \cite{arimondo96} with
\begin{eqnarray}
\rho_{ii}(v, t) &=& W(v) R_{ii}(t), \nonumber \\
\tilde{\rho}_{ij}(v, t) &=& W(v) R_{ij}(t), ~~~ i \ne j . \label{thermal}
\end{eqnarray}
Here $R_{ij}(t)$s are dimensionless, and $\sum_i R_{ii} (t) = 1$, for a closed system without dissipation. The simplified assumption
(\ref{thermal}) would imply that the effect of the VCCs is to bring about a complete redistribution of the population over all the velocity
classes such that the inhomogeneous media is similar to a homogeneous one but with a width given by the inhomogeneous Doppler-broadened
width. Hence, the problem effectively reduces to that of a homogeneous system with a Doppler-broadened pumping rate.

\subsubsection{Decoherence by VCC}

However, VCCs affect populations and coherences in a slightly different manner. Though there is a redistribution of the atomic population
over the entire inhomogeneous width, VCCs can still lead to a decoherence of the Raman coherence. In order to take account of any
depolarization because of collisions that leads to a decoherence of the prepared EIT medium, we additionally modify the collision kernel in
Eq.\,(\ref{eq11}) for the lower-state coherence \cite{arimondo96} with strong thermalizing VCCs by inserting a depolarization ratio $\alpha$
as
\begin{equation}
K_{1}(v' \rightarrow v) = \alpha~ \Gamma_{1 \mathrm{VCC}}~ W(v),  \label{nonconserv}
\end{equation}
with $\alpha \leq 1$ from Eq.\,(\ref{eq7}). We define the parameter $\eta \equiv 1 - \alpha$ as a deviation from complete coherence
preservation. $1 > \eta > 0$ would imply a loss of coherence by VCCs, and hence a loss of coherently-prepared, dark-state atoms from the
Maxwell-Boltzmann distributed system. $\eta = 0$ corresponds to a complete redistribution of population over the Doppler width by VCC,
without any loss of polarization.

The VCC contribution gets added to the transit time decay (and hence the Raman coherence relaxation) rate as $\eta \Gamma_{\mathrm{1VCC}}$
(with $\eta\approx$ 0). The populations in the lower levels are aided by the influx of fresh atoms at the dynamic transit rate (a part of
which has no contribution from the VCC), yielding the inhomogeneity in the density matrix equations in the steady state, necessary for
non-trivial solutions.

The steady-state solutions of Eqs.\,(\ref{eq14})-(\ref{eq19}) with particular combinations such as $\rho_{aa}$, $(\rho_{bb} \pm \rho_{cc})$
and $(\tilde{\rho}_{cb} \pm \tilde{\rho}_{bc})$ are considered. Using Eqs.\,({\ref{thermal}) and ({\ref{nonconserv}}), and integrating the
relevant equations over velocity, we obtain the following:
\begin{eqnarray}
- [\Gamma_0 + \Gamma_{\mathrm{t}} ] R_{aa} - [\Gamma_{\mathrm{X4}} (\Delta + \delta_{\mathrm{R}})] (R_{aa} -R_{bb}) - \Gamma_{\mathrm{X3}} (\Delta) (R_{aa} - R_{cc})
+ [\Gamma_{\mathrm{X1}} (\Delta + \delta_{\mathrm{R}}) ~~~~~~~~\nonumber \\ + \Gamma_{\mathrm{X1}} (\Delta)] \left( \frac{R_{cb} + R_{bc}}{2}\right) +i [\Gamma_{\mathrm{Y1}} (\Delta + \delta_{\mathrm{R}}) - \Gamma_{\mathrm{Y1}} (\Delta)] \left( \frac{R_{cb} - R_{bc}}{2} \right) = 0,~~~~~~~~~~~~~~~~~~~~~ \label{eq20} \\
\Gamma_0 R_{aa} - \Gamma_{\mathrm{t}} (R_{bb}+R_{cc}) + \Gamma_{\mathrm{t}} + [\Gamma_{\mathrm{X4}} (\Delta + \delta_{\mathrm{R}})] (R_{aa} - R_{bb}) + \Gamma_{\mathrm{X3}} (\Delta) (R_{aa} - R_ {cc}) ~~~~~~~~\nonumber \\
- [\Gamma_{\mathrm{X1}} (\Delta + \delta_{\mathrm{R}}) + \Gamma_{\mathrm{X1}} (\Delta)] \left( \frac{R_{cb} + R_{bc}}{2}\right)- i [\Gamma_{\mathrm{Y1}} (\Delta + \delta_{\mathrm{R}}) - \Gamma_{\mathrm{Y1}} (\Delta)] \left( \frac{R_{cb} - R_{bc}}{2} \right) = 0, \label{eq21} \\
- [ (1- \beta) \Gamma_{\mathrm{t}} ] (R_{bb}-R_{cc}) + [\Gamma_{\mathrm{X4}} (\Delta + \delta_{\mathrm{R}})] (R_{aa} - R_{bb}) - \Gamma_{\mathrm{X3}} (\Delta) (R_{aa} - R_{cc}) ~~~~~~~~~~~~~~~~\nonumber \\
- [\Gamma_{\mathrm{X1}} (\Delta + \delta_{\mathrm{R}}) - \Gamma_{\mathrm{X1}} (\Delta)] \left( \frac{R_{cb} + R_{bc}}{2} \right) - i [\Gamma_{\mathrm{Y1}} (\Delta + \delta_{\mathrm{R}}) + \Gamma_{\mathrm{Y1}} (\Delta)] \left( \frac{R_{cb} - R_{bc}}{2} \right) = 0, \label{eq22} \\
- \left[ 2 \Gamma_{\mathrm{R}} + 2 \eta~ \Gamma_{1\mathrm{VCC}} + \Gamma_{\mathrm{X3}} (\Delta + \delta_{\mathrm{R}}) + \Gamma_{\mathrm{X4}} (\Delta) \right] \left( \frac{R_{cb} + R_{bc}}{2} \right)
+ i \Big[ 2 \delta_{\mathrm{R}} - \Gamma_{\mathrm{Y3}} (\Delta + \delta_{\mathrm{R}}) \nonumber \\ + \Gamma_{\mathrm{Y4}} (\Delta) \Big] \left( \frac{R_{cb} - R_{bc}}{2} \right)
+ \Gamma_{\mathrm{X1}} (\Delta + \delta_{\mathrm{R}}) (R_{aa} - R_{bb}) + \Gamma_{\mathrm{X1}} (\Delta) (R_{aa} - R_{cc}) = 0, ~~~~~~~ \label{23} \\
- \left[ 2 \Gamma_{\mathrm{R}} + 2 \eta~ \Gamma_{1\mathrm{VCC}} + \Gamma_{\mathrm{X3}} (\Delta + \delta_{\mathrm{R}}) + \Gamma_{\mathrm{X4}} (\Delta) \right] \left( \frac{R_{cb} - R_{bc}}{2} \right)
+ i \Big[ 2 \delta_{\mathrm{R}} - \Gamma_{\mathrm{Y3}} (\Delta + \delta_{\mathrm{R}})  \nonumber \\ + \Gamma_{\mathrm{Y4}} (\Delta) \Big] \left( \frac{R_{cb} + R_{bc}}{2}\right)
+ \Gamma_{\mathrm{Y1}} (\Delta + \delta_{\mathrm{R}}) (R_{aa} - R_{bb}) - \Gamma_{\mathrm{Y1}} (\Delta) (R_{aa} - R_{cc}) = 0. ~~~~~~ \label{24}
\end{eqnarray}
Here the different Doppler-broadened rates have the following forms:
\begin{eqnarray}
\Gamma_{\mathrm{X1}} (\Delta) &=& \left( \frac{\Omega_{\mathrm{C}} \Omega_{\mathrm{P}}}{2} \right) ~V_{\mathrm{X}} (\Delta), \label{25} \\
\Gamma_{\mathrm{X2}} (\Delta) &=& \left( \frac{\Omega_{\mathrm{C}}^2 + \Omega_{\mathrm{P}}^2}{2} \right) ~V_{\mathrm{X}} (\Delta), \label{26} \\
\Gamma_{\mathrm{X3}} (\Delta) &=& \left( \frac{\Omega_{\mathrm{C}}^2}{2} \right) ~V_{\mathrm{X}} (\Delta), \label{27} \\
\Gamma_{\mathrm{X4}} (\Delta) &=& \left( \frac{\Omega_{\mathrm{P}}^2}{2} \right) ~V_{\mathrm{X}} (\Delta), \label{28} \\
\Gamma_{\mathrm{Y1}} (\Delta) &=& \left( \frac{\Omega_{\mathrm{C}} \Omega_{\mathrm{P}}}{2} \right) ~V_{\mathrm{Y}} (\Delta), \label{29} \\
\Gamma_{\mathrm{Y2}} (\Delta) &=& \left( \frac{\Omega_{\mathrm{C}}^2 - \Omega_{\mathrm{P}}^2}{2} \right) ~V_{\mathrm{Y}} (\Delta), \label{30} \\
\Gamma_{\mathrm{Y3}} (\Delta) &=& \left( \frac{\Omega_{\mathrm{C}}^2}{2} \right) ~V_{\mathrm{Y}} (\Delta), \label{31} \\
\Gamma_{\mathrm{Y4}} (\Delta) &=& \left( \frac{\Omega_{\mathrm{P}}^2}{2} \right) ~V_{\mathrm{Y}} (\Delta), \label{32}
\end{eqnarray}
where
\begin{eqnarray}
V_{\mathrm{X}} (\Delta) &=& \frac{\Gamma /2}{\sqrt{\pi} u} \int_{-\infty}^{\infty} \frac{e^{-{v^2}/{u^2}} ~dv}{(\Gamma /2)^2+(\Delta - k v)^2} = \frac{\sqrt{\pi}}{k u} \left[ 1- \mbox{Erf} (p) \right] e^{[p^2 - q^2(\Delta)]} \cos[2pq(\Delta)], \label{33} \\
V_{\mathrm{Y}} (\Delta) &=& \frac{1}{\sqrt{\pi} u} \int_{-\infty}^{\infty} \frac{(\Delta - k v) e^{-{v^2}/{u^2}} ~dv}{(\Gamma /2)^2+{(\Delta - k v)}^2}, \label{34}
\end{eqnarray}
with $p = \Gamma /(2k u)$ and $q(\Delta) = \Delta /(k u)$.

From the above set of equations using these Doppler-broadened rates, we finally obtain the steady-state value of $R_{ab}$ which yields $\rho_{ab}$.
\begin{eqnarray}
\mbox{Re} [R_{ab}] &=& -\frac{\Gamma_{\mathrm{X3}} (\Delta + \delta_{\mathrm{R}})}{\Omega_{\mathrm{C}}} \mbox{Im} [R_{cb}] - \frac{\Gamma_{\mathrm{Y3}} (\Delta + \delta_{\mathrm{R}})}{\Omega_{\mathrm{C}}} \mbox{Re} [R_{cb}] + \frac{\Gamma_{\mathrm{Y4}} (\Delta + \delta_{\mathrm{R}})}{\Omega_{\mathrm{P}}} (R_{aa} - R_{bb}), \\
\mbox{Im} [R_{ab}] &=& \frac{\Gamma_{\mathrm{X3}} (\Delta + \delta_{\mathrm{R}})}{\Omega_{\mathrm{C}}} \mbox{Re} [R_{cb}] - \frac{\Gamma_{\mathrm{Y3}} (\Delta + \delta_{\mathrm{R}})}{\Omega_{\mathrm{C}}} \mbox{Im} [R_{cb}] - \frac{\Gamma_{\mathrm{X4}} (\Delta + \delta_{\mathrm{R}})}{\Omega_{\mathrm{P}}} (R_{aa} - R_{bb}),
\end{eqnarray}
where $\mbox{Re} [R_{cb}] = \frac{R_{cb} + R_{bc}}{2}$ and $\mbox{Im} [R_{cb}] = \frac{R_{cb} - R_{bc}}{2}$.

The susceptibility for the Doppler-broadened medium is given as:
\begin{equation}
\chi = \mbox{A} ~ R_{ab},
\end{equation}
where $\mbox{A}$ is the normalization constant, which is obtainable from the initial probe transmission at resonance in the absence of the coupling field.

\section{Results}

We now test the validity of our theory by comparing our predictions against known experimental results in hot vapors, viz. in a He* cell
\cite{goldfarb08}. For He*, collisions are expected to play a favorable role through five different effects: (i) VCCs enable one to span the
entire Doppler profile quickly and efficiently during optically pumping; (ii) collisions aid in the feeding distortion in pumping back more
atoms in $\vert b \rangle$ compared to that in $\vert c \rangle$ when they are entering the beam; (iii) collisions increase the transit time
of the atoms through the beam and hence the Raman coherence lifetime \cite{fitzsimmons68};  and (iv) this is possible because collisions
involving He atoms in the zero spin and angular momentum ground state do not depolarize the colliding He* \cite{phelps55}; and (v) Penning
ionization (PI) among identically polarized He* atoms is almost forbidden \cite{shlyapnikov94}.

We probe the general dependence of various features of interest -- Doppler-averaged Fano-like transmission profiles, the variation of the
EIT width, the peak transmission and the group delay with the coupling intensity -- on the following system characteristics: the VCC
parameter $\eta$, the unequal atomic influx parameter $\beta$, the Raman decoherence rate $\Gamma_{\mathrm{R}}$, and the initial
transmission $T_0$ which carries the information about the number density of the participating atoms. In our model, the effect of collisions
is incorporated through collisional dephasing of both the optical and Raman coherences given in Eqs.\,(\ref{optcoh}) and (\ref{Ramancoh}),
the collision-induced diffusive (as opposed to ballistic) transit rate $\Gamma_{\mathrm{t}}$ of the atoms, a complete redistribution of the
atomic population over all velocity classes leading to a velocity-independent optical pumping under rapid VCC coverage, and the VCC
decoherence rate $\Gamma_{1 ~\mathrm{VCC}}$ entering when the VCC parameter $\eta \neq 0$. For the He* system, we find that the decoherence
due to VCC is very small. As $\eta$ is a deviation from the complete polarization-conservation by VCC, it should not depend on the beam
size. We choose a value of $\eta = 10^{-4}$, which helps us to obtain good fits to the measured evolutions with the coupling intensity for
each of EIT width, peak transmission and group delay, for all the measured beam diameters, with $\Gamma_{1~ \mathrm{VCC}} = 10^7
~\mathrm{s}^{-1}$, which is simply the number of collisions per second as mentioned earlier. The excess incoming rate, $\beta
~\frac{\Gamma_\mathrm{t}}{2} \left( \rho_{bb}(v) - \rho_{cc}(v) \right)$ to state $\vert b \rangle$ over $\vert c \rangle$ also has a
distinct impact on the various features. In particular, the parameter $\beta$ = 0.1 gives satisfactory results in comparing the theoretical
results with the experimentally measured evolutions. With the precisions of our measurements in He*, a constant value of $\beta$ is found to
fit well for all beam diameters. The choice of the best-fit values of the parameters $\eta$ and $\beta$ for the He* system has been
elaborated using the group delay plots in subsection D below.

There was a slight over-estimation of the initial transmission $T_0$ for the EIT experiments in He*, reported in Ref.\,\cite{goldfarb08}
directly from the measured values, which included an effect of partial saturation by the probe power. This has been adjusted in the present
paper.

\subsection{Doppler-broadened Fano-like profiles}

When the coupling beam frequency is no longer at the center of the Doppler profile, transmitted intensity profiles become asymmetric. This
is similar to the Fano profiles obtained in the case of EIT in a homogeneously-broadened medium and which have been shown to be due to
interferences between a direct process and stimulated Raman scattering in the overall transition probability \cite{lounis92, wong04}.
However, here, these profiles are modified by the fact that they have to be convoluted with the inhomogeneous Doppler profile. Figure
\ref{fig2} shows the evolution of the transmission versus Raman detuning $\delta_{\mathrm{R}}$ for different values of the detuning $\Delta$
of the coupling beam with respect to the center of the Doppler profile. On the left-hand side, the measured values through a He* cell with
initial transmission at resonance $T_0$ = 0.56 for a beam diameter of 1.5 cm, a coupling power of 11 mW and a probe power of 140 $\mu$W are
reproduced from Ref.\,\cite{goldfarb08}. The base of the symmetric (black) curve at resonance corresponds to a transmission higher than
0.56, because of partial saturation of the medium by the probe intensity.

\begin{figure}[htbp]
\scalebox{0.43}{\includegraphics{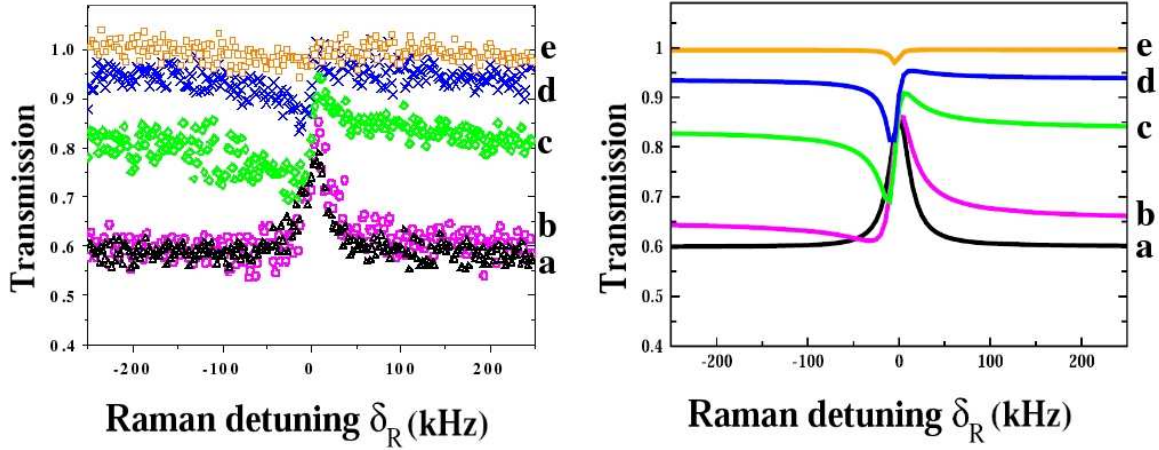}}
\caption{\label{fig2} Evolution of the transmission versus Raman detuning $\delta_{\mathrm{R}}$ for different values of the detuning $\Delta$ of the coupling beam with respect to the center of the Doppler profile: on the left-hand side, the measured values through a He* cell with $T_0$ = 0.56 for a beam diameter of 1.5 cm, a coupling power of 11 mW and a probe power of 140 $\mu$W with (a) $\Delta = 0$ (black, triangles), (b) $\Delta = 0.4$ GHz (magenta, open circles), (c) $\Delta = 1.0$ GHz (green, diamonds), (d) $\Delta = 1.4$ GHz (blue, crosses), and (e) $\Delta = 2.1$ GHz (orange, open squares), are reproduced from Ref.\,\cite{goldfarb08}. On the
right-hand side, theoretical evolution of the transmission versus Raman detuning $\delta_{\mathrm{R}}$, corresponding to those obtained experimentally are shown with (a) $\Delta = 0$ (black), (b) $\Delta = 0.4$ GHz (magenta), (c) $\Delta = 1.0$ GHz (green), (d) $\Delta = 1.4$ GHz (blue), and (e) $\Delta = 2.1$ GHz (orange), using $\Gamma_{\mathrm{R}}/2\pi$ = 4.3 kHz, $\Gamma_{\mathrm{t}}/2\pi$ = 0.41 kHz, $\eta = 10^{-4}$, and $\beta$ = 0.1.}
\end{figure}

Theoretically, the transmission profiles are generated from the imaginary part of the susceptibility at different detunings $\Delta$ of the
coupling beam with respect to the center of the Doppler profile mentioned above. The corresponding plots on the right-hand side of
Fig.\,\ref{fig2} are obtained for a beam diameter of 1.5 cm corresponding to different values of $\Delta$, with the same initial
transmission, coupling and probe powers as in the experiment, using $\Gamma_{\mathrm{R}}/2\pi$ = 4.3 kHz, $\Gamma_{\mathrm{t}}/2\pi$ = 0.41
kHz, $\eta = 10^{-4}$, and $\beta$ = 0.1. The transmission profiles at all values of $\Delta$ increases slightly with an increase in
$\beta$. This is also seen in the behavior of peak transmission shown in subsection C. This has been checked for a range of $\beta$ from 0
to 0.1. There is a distinct and sensitive dependence on $\eta$, as the tip of each profile decreases slightly with an increase of $\eta$,
and the effect is the largest on the resonant profile. As a result, there is also a change in the relative placement of the profiles. This
has been checked for a range of values of $\eta$ from 10$^{-3}$ to 0. The reason for this is simple. Since a deviation of the value of
$\eta$ from 0 indicates a loss of coherence, the transmission at resonance will be greater for $\eta$ closer to 0, because the system is
more coherent as $\eta$ approaches 0. Additionally, the effect will be more pronounced when the atomic system is in resonance with the two
fields, which is the condition for a perfect EIT. We thus find that the predictions from our model indeed agree very well with the
experimental results.

\subsection{EIT width}

The mechanism of decoherence in EIT in a room-temperature gas can be probed by measuring the width of the EIT resonance as a function of the
coupling beam intensity. As mentioned in the Introduction, the theoretical treatment of EIT in Doppler-broadened gases \cite{javan02},
assuming the population exchange between the lower levels to be the main source of decoherence, predicts a {\it nonlinear} dependence for
weak coupling powers: the EIT width is expected to evolve with the coupling beam Rabi frequency $\Omega_{\mathrm{C}}$ \cite{Rabi} according
to:
\begin{equation}
\Gamma_{\mathrm{EIT}}\simeq \frac{\Omega_{\mathrm{C}}^2}{4\delta_{\mathrm{eff}}}\;, \label{javan}
\end{equation}
where $\delta_{\mathrm{eff}}$ gives the effective width over which the atoms are pumped into the probe ground state for a fixed value of
$\Omega_{\mathrm{C}}$. In the case when $\Omega_{\mathrm{C}} \ll \Omega_{\mathrm{inhom}} =
2\sqrt{2\Gamma_{\mathrm{R}}/\Gamma}W_{\mathrm{D}}$, $\delta_{\mathrm{eff}}$ is dependent on $\Omega_{\mathrm{C}}$ and $\Gamma_{\mathrm{R}}$:
$\delta_{\mathrm{eff}} = \Omega_{\mathrm{C}}\sqrt{\Gamma_0 /8\Gamma_{\mathrm{R}}}$, and the coherent population trapping is shown to be
velocity selective, i.e., it occurs only for those atoms whose frequencies are close to resonance with the coupling field. In the opposite
regime, when $\Omega_{\mathrm{C}} \gg \Omega_{\mathrm{inhom}}$, Ref.\,\cite{javan02} predicts that $\delta_{\mathrm{eff}} = W_{\mathrm{D}}$,
i.e., in the limit of very large intensity, it is reduced to the power-broadening case.

In contrast, in Ref.\,\cite{lvovsky06} it is assumed that all the atoms across the Doppler profile are initially optically pumped by the
coupling beam to the probe ground level, and the decoherence in the lower states is caused by pure dephasing, and not population exchange.
Then the response of the medium up to first order in probe field yields the following {\it linear} dependence of the EIT linewidth on the
coupling beam intensity:
\begin{equation}
\Gamma_{\mathrm{EIT}}=2\Gamma_{\mathrm{R}}+\frac{\Omega_{\mathrm{C}}^2}{2W_{\mathrm{D}}+\Gamma}\;. \label{lvovsky}
\end{equation}

An example of the evolution of the EIT width versus the intensity of the coupling beam is displayed in Fig.\,\ref{fig3}(a). In the
experiment with He* \cite{goldfarb08}, the logarithm of the transmitted intensity is calculated and its full width at half maximum (FWHM) is
then measured, in order to determine precisely the width of the susceptibility $\chi$ of the medium. The measured sub-natural ($< \Gamma$)
widths for a beam diameter of 1.5 cm, a probe power of 100 $\mu$W and an initial transmission $T_0$ = 0.46 are shown as dots. Theoretically,
the imaginary part of the susceptibility is fitted with a Lorentzian to obtain the FWHM corresponding to a particular coupling intensity.
The continuous line in Fig.\,\ref{fig3} is the best fit from our model with the same parameters as in the experiment, using
$\Gamma_{\mathrm{R}}/2 \pi$ = 4.3 kHz, $\Gamma_{\mathrm{t}}/2 \pi$ = 0.41 kHz, $\eta = 10^{-4}$, and $\beta$ = 0.1. The EIT width
$\Gamma_{\mathrm{EIT}}$ is seen to evolve linearly with the coupling beam intensity, i.e., quadratically with the coupling beam Rabi
frequency $\Omega_{\mathrm{C}}$.

\begin{figure}[htbp]
\scalebox{0.45}{\includegraphics{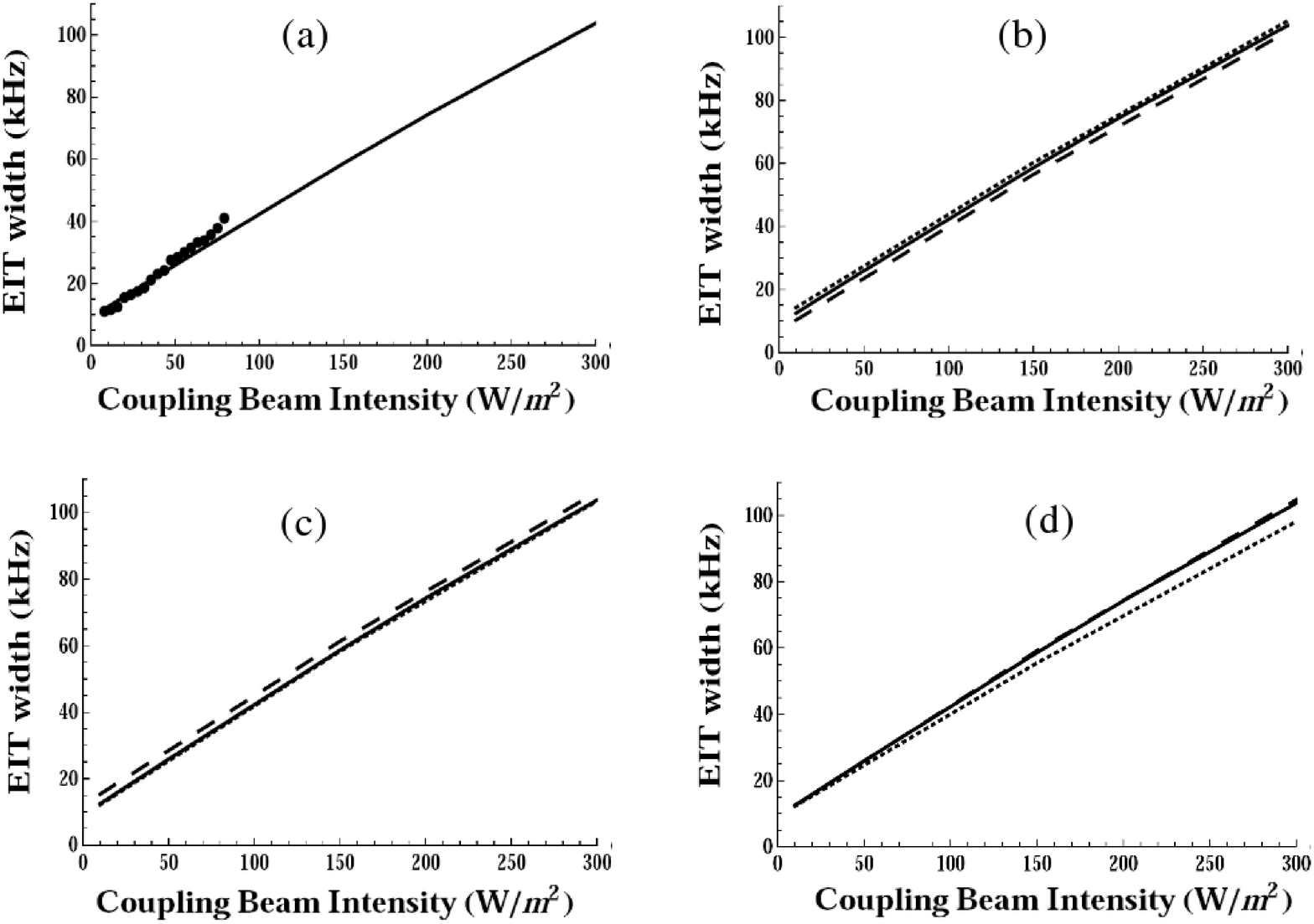}} \caption{\label{fig3} Evolution of the EIT window width versus coupling beam
intensity for a beam diameter of 1.5 cm with a probe power of 100 $\mu$W: (a) the experimentally measured values for $T_0$ = 0.46,
reproduced from Ref.\,\cite{goldfarb08}, are shown as dots, along with the theoretical best fit (continuous line) from our model using $\eta
= 10^{-4}$, $\beta$ = 0.1, $\Gamma_{\mathrm{R}}/2\pi$ = 4.3 kHz and $\Gamma_{\mathrm{t}} /2\pi$ = 0.41 kHz. (b) Predictions from our model
for different values of the Raman decoherence rate: (i) $\Gamma_{\mathrm{R}}/ 2\pi$ = 3.2 kHz (dashed), (ii) $\Gamma_{\mathrm{R}}/2\pi$ =
4.3 kHz (continuous), and (iii) $\Gamma_{\mathrm{R}}/2\pi$ = 5.2 kHz (dotted), with the rest of the parameters the same as in (a). (c)
Predictions from our model for different values of the VCC parameter: (i) $\eta = 10^{-3}$ (dashed), (ii) $\eta = 10^{-4}$ (continuous), and
(iii) $\eta = 0$ (dotted), with the rest of the parameters the same as in (a). (d) Predictions from our model for different values of the
optical coherence decay rate: (i) $\Gamma /2\pi$ = 0.1 MHz (dashed), (ii) $\Gamma /2\pi$ = 22.3 MHz (continuous), and (iii) $\Gamma /2\pi$ =
150 MHz (dotted). The continuous line is the same in all the figures.}
\end{figure}

With the experimental parameters ($\Gamma=1.4\times10^8\,\mathrm{s}^{-1}$ at 1 Torr, $\Gamma_{\mathrm{R}}=10^4-10^5\,\mathrm{s}^{-1}$,
$W_{\mathrm{D}}/2\pi=0.9$ GHz) \cite{goldfarb08}, one obtains $10^8\,\mathrm{rad/s} \leq \Omega_{\mathrm{inhom}} \leq 4\times
10^8\,\mathrm{rad/s}$. Since the maximum Rabi frequencies $\Omega_{\mathrm{C}}$ used in the experiment are smaller than $5\times
10^7\,\mathrm{rad/s}$, the measurements are in the first regime of \cite{javan02}, where $\Omega_{\mathrm{C}} \ll \Omega_{\mathrm{inhom}}$.
Thus $\Gamma_{\mathrm{EIT}}$ is predicted to evolve linearly with $\Omega_{\mathrm{C}}$ \cite{javan02}, with a slope depending on
$\Gamma_{\mathrm{R}}$, but these predictions are violated in the experiment. If we use Eq.\,(\ref{javan}) to fit the linear evolution of
$\Gamma_{\mathrm{EIT}}$ versus the coupling intensity, we obtain $\delta_{\mathrm{eff}}/2\pi =\ 0.5\ \mathrm{GHz}$, which is of the same
order of magnitude as $W_{\mathrm{D}}/2\pi$, showing that a major part of the Doppler profile takes part in the EIT process.

In our model we achieve the straight line feature of $\Gamma_{\mathrm{EIT}}$ versus coupling intensity as in Ref.\,\cite{lvovsky06}, but
without any assumptions of special initial conditions. Moreover, if the data are fitted to the straight line given by Eq.\,(\ref{lvovsky}),
the resulting estimate of $\Gamma_{\mathrm{R}}/2\pi$ does not reproduce accurately the measured evolutions of the peak-transmission and
delay \cite{goldfarb08,f}. Our present model fits the straight line data for atomic $^{87}$Rb vapor at temperatures 60-100$^{\circ}$C given
in Ref.\,\cite{lvovsky06}, with appropriate values of the parameters for the optically thick system.

Keeping all other parameters constant, a change in the initial transmission $T_0$, and hence in the number density of the participating
atoms in the cell, does not affect the evolution of the EIT width with coupling intensity in our model. Likewise, a change in the unequal
feeding parameter $\beta$ does not produce any visible change in the EIT width. This has been checked for a range of values of $\beta =$ 0
to 0.5. This is understandable as these two parameters do not affect the Raman coherence lifetime but affect the pumping efficiency (e.g.
the fraction of atoms that participate in the EIT phenomenon).}

Figure \ref{fig3}(b) shows the variation of the EIT width with the coupling intensity from our model for different values of the Raman
decoherence rate $\Gamma_{\mathrm{R}}$ around the best-fit value, keeping $\eta$, $\beta$ and $T_0$ constant. It clearly shows that the
slope of the evolution is the same for different values of $\Gamma_{\mathrm{R}}$. Note that different laser beam diameters would lead to
different transit times of the atoms through the beam, and hence different $\Gamma_{\mathrm{t}}$. The motion of the atoms through the beam
is assumed to be diffusive, as stated previously, because of the large number of collisions they undergo. The Raman decoherence rate
$\Gamma_{\mathrm{R}}$ given by Eq.\,(15) thus contains a variable $\Gamma_{\mathrm{t}}$ plus other terms. For the He* measurements, we
estimate that $(\Gamma_{\mathrm{R}}-\Gamma_{\mathrm{t}})/2 \pi \approx$ 4 kHz for a low pressure of 1 Torr and an ambient magnetic field
inhomogeneity. This is obtained by estimating first $\Gamma_{\mathrm{t}}$ for a particular beam diameter, and then $\Gamma_{\mathrm{R}}$
from the best fit of (primarily) the EIT-width data. In the experiment, the beam diameters are generally not precisely known, because the
beam emerging from the acousto-optic modulators is not perfectly Gaussian. This affects the precise determination of the transit times as
well as beam intensities (the probe and coupling powers measured before the He* cell were, of course, their average values) and Rabi
frequencies. We have, however, checked that the experimental data for different beam sizes yield the $\Gamma_{\mathrm{t}}$ values following
the diffusive transit scenario.

Figure \ref{fig3}(c) shows the variation for different values of the VCC parameter $\eta$. The slope of the evolution is the same for
different values of $\alpha$ but the intercept increases with an decrease of $\eta$. From Figs.\,\ref{fig4}(b) and (c), the width-intercept
at $\Omega_{\mathrm{C}}$ = 0 are seen to depend on the parameters $\Gamma_{\mathrm{R}}$ and $\eta$, yielding a narrower resonance for a
lower $\Gamma_{\mathrm{R}}$ or a lower $\eta$. The inset shows that the measured data fit better with $\eta \neq 0$ and $\beta \neq 0$.
However, the lines do not fit Eq.\,(\ref{lvovsky}), with $\Gamma_{\mathrm{R}}$ of Ref.\,\cite{lvovsky06} simply replaced here by
$\Gamma_{\mathrm{R}} + \eta~ \Gamma_{1~ \mathrm{VCC}}$. As pointed out earlier, in the density-matrix equations, the VCC contribution gets
added to the transit time decay (and hence the Raman coherence relaxation) rate as $\eta~ \Gamma_{1~ \mathrm{VCC}}$, except in the
inhomogeneous term signifying atomic influx in the populations in the lower levels at the transit rate $\Gamma_{\mathrm{t}}$ which has no
contribution from the VCC. It is therefore not surprising that the net effect of $\eta$ cannot be taken care of by $\Gamma_{\mathrm{t}}$ or
$\Gamma_{\mathrm{R}}$.

Figure \ref{fig3}(d) shows the variation of the EIT width with the coupling intensity from our model for different values of the optical
coherence decay rate $\Gamma$. The slope of the line changes inversely with changes in the values of $W_{\mathrm{D}}$ and $\Gamma$.

\subsection{Peak transmission}

\begin{figure}[htbp]
\scalebox{0.45}{\includegraphics{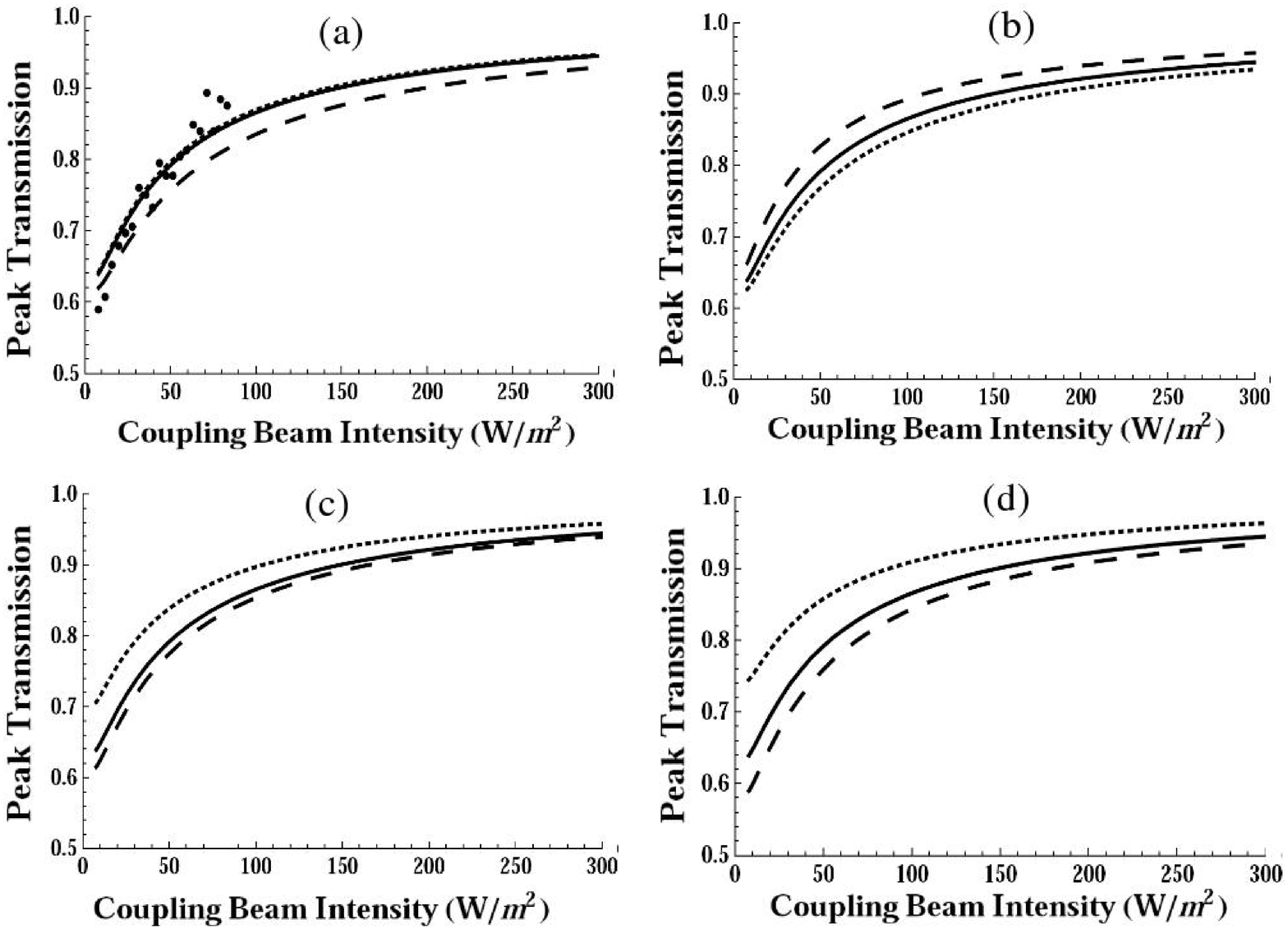}} \caption{\label{fig4} Evolution of the cell transmission versus coupling beam
intensity for a beam diameter of 1.5 cm with a probe power of 70 $\mu$W and $T_0$ = 0.46: (a) the experimentally measured values through a
He* cell, reproduced from Ref.\,\cite{goldfarb08}, are shown as dots, along with the theoretical best fit (continuous curve) from our model
using $\eta = 10^{-4}$, $\beta$ = 0.1, $\Gamma_{\mathrm{R}}/2\pi$ = 4.3 kHz and $\Gamma_{\mathrm{t}}/2\pi$ = 0.41 kHz. Also shown are the
predictions from our model for different values of the VCC parameter: $\eta = 10^{-3}$ (dashed), and $\eta$ = 0 (dotted). (b) Predictions
from our model for different values of the Raman decoherence rate: (i) $\Gamma_{\mathrm{R}}/ 2\pi$ = 3.2 kHz (dashed), (ii)
$\Gamma_{\mathrm{R}}/2\pi$ = 4.3 kHz (continuous), and (iii) $\Gamma_{\mathrm{R}}/2\pi$ = 5.2 kHz (dotted), with the rest of the parameters
the same as in the continuous fit in (a). (c) Predictions from our model for different values of the unequal feeding parameter $\beta$: (i)
$\beta$ = 0 (dashed), (ii) $\beta$ = 0.1 (continuous), and (iii) $\beta$ = 0.5 (dotted), with the rest of the parameters the same as in the
continuous fit in (a). (d) Predictions from our model for different values of the initial transmission: (i) $T_0$ = 0.4 (dashed), (ii) $T_0$
= 0.46 (continuous), and (iii) $T_0$ = 0.6 (dotted), with the rest of the parameters the same as in the continuous fit in (a). The
continuous curve is the same in all the figures.}
\end{figure}

We next consider the maximum probe transmission at resonance ($\Delta$ = 0) corresponding to different coupling intensities. Figure
\ref{fig4}(a) shows the evolution of the cell transmission versus coupling beam intensity. The measured values (dots) through a He* cell for
a beam diameter of 1.5 cm are reproduced from Ref.\,\cite{goldfarb08}. The corresponding continuous curve is obtained from our theory by
calculating the transmission at two-photon resonance $\delta_{\mathrm{R}} = 0$, with $\eta = 10^{-4}$, $\beta$ = 0.1 at the same initial
transmission and probe power as in the experiment: probe power of 70 $\mu$W, $T_0$ = 0.46, $\Gamma_{\mathrm{R}}/2\pi$ = 4.3 kHz, and
$\Gamma_{\mathrm{t}}/2 \pi$ = 0.41 kHz.

Using the same hypotheses as for the derivation of Eq.\,(\ref{lvovsky}) \cite{lvovsky06}, the peak transmission is predicted to evolve as \cite{goldfarb08}:
\begin{equation}
\ln(T) = \frac{\ln(T_0)}{1+\frac{\Omega_{\mathrm{C}}^2}{2\Gamma_{\mathrm{R}}(2W_{\mathrm{D}}+\Gamma)}}\;. \label{lvovsky_trans}
\end{equation}
It is seen that there is a better agreement of the experimental result with our present theory than what was obtained using
Eq.\,(\ref{lvovsky_trans}) in Ref.\,\cite{goldfarb08} with the value of $\Gamma_{\mathrm{R}}$ deduced from the EIT-width data, in spite of
the fact that we do not assume $\rho_{bb}(t=0)$ = 1, i.e., a system perfectly prepared by optical pumping at $t$ = 0 as in
Ref.\,\cite{lvovsky06}, is redundant here in the presence of rapid VCC coverage.

Figure \ref{fig4}(a) also shows the variation of the peak transmission with the coupling intensity from our model for different values of
the VCC parameter $\eta$. The dotted and the dashed plots correspond to $\eta$ = 0 and $10^{-3}$, respectively. It can be thus inferred that
the peak transmission increases with a decrease in $\eta$. This also has support from the effect of $\eta$ on the Doppler-broadened profiles
in Fig.\,\ref{fig2} for the same reason. The best-looking fit for $\eta = 10^{-4}$ emphasizes that the depolarization due to VCCs is a small
effect, as expected from the work of Shlyapnikov \textit{et al.} \cite{shlyapnikov94}.

Figure \ref{fig4}(b) shows the variation of the peak transmission with the coupling intensity from our model for different values of the
Raman decoherence rate $\Gamma_{\mathrm{R}}$ around the best-fit value, when $\eta$, $\beta$ and $T_0$ are kept constant \cite{comment}. It
is clear that a lower decoherence rate $\Gamma_{\mathrm{R}}$ leads to a more coherent system, and thus $\Gamma_{\mathrm{R}}/2\pi$ = 3.2 kHz
(dashed curve) yields the highest peak transmission compared to the rest shown in this plot. Thus the evolutions from our model are
consistent with our physical understanding. Also note that Figs.\,\ref{fig4}(a) and (b) show that an increase in $\Gamma_{\mathrm{R}}$ has a
similar effect as a decrease in $\eta$.

Figure \ref{fig4}(c) shows the same for different values of the unequal feeding parameter $\beta$ from the theory. The effect of $\beta$
here is clearly very distinct. As $\beta$ increases, there are more atoms entering the laser beam in the dark state $\vert b \rangle$. Thus,
the peak transmission at two-photon resonance will also increase as the transparency is more when more atoms can participate in the EIT
phenomenon.

Figure \ref{fig4}(d) shows the effect of different values of the initial transmission $T_0$. It is obvious that a higher initial
transmission $T_0$ would lead to a higher peak transmission for non-zero coupling intensities.

\subsection{Group delay}

\begin{figure}[htbp]
\scalebox{0.43}{\includegraphics{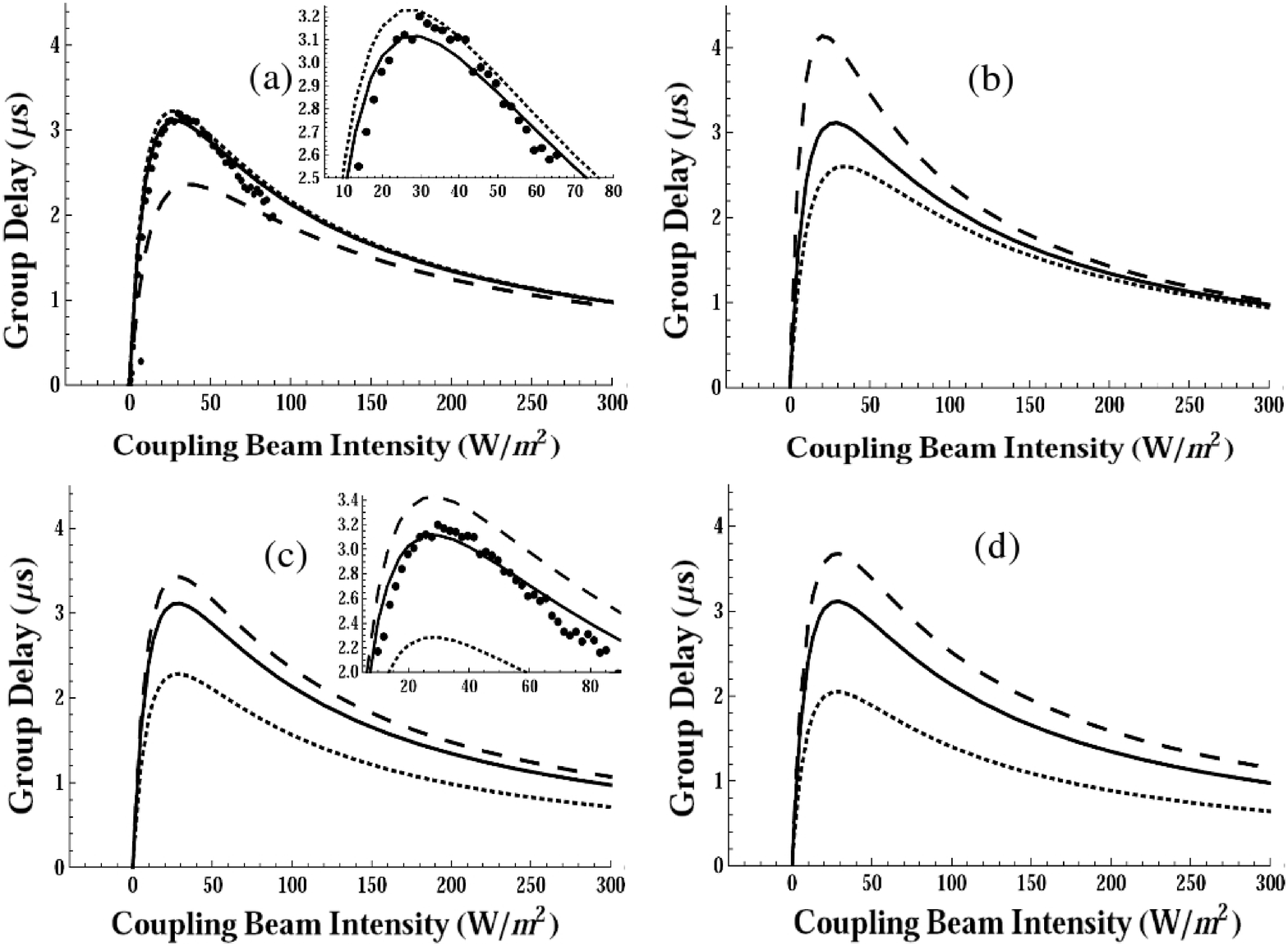}} \caption{\label{fig5} Evolution of the group delay with the coupling
intensity for a beam diameter of 1.5 cm, with a probe power of 35 $\mu$W and $T_0$ = 0.46: (a) the experimentally measured values through a
He* cell, reproduced from Ref.\,\cite{goldfarb08}, are shown as dots, along with the theoretical best fit (continuous curve) from our model
using $\eta = 10^{-4}$, $\beta$ = 0.1, $\Gamma_{\mathrm{R}}/2\pi$ = 4.3 kHz and $\Gamma_{\mathrm{t}}/2\pi$ = 0.41 kHz. Also shown are the
predicted variations from our model for different values of the VCC parameter: $\eta = 10^{-3}$ (dashed), and $\eta = 0$ (dotted). (b)
Predictions from our model for different values of the Raman decoherence rate: (i) $\Gamma_{\mathrm{R}}/ 2\pi$ = 3.2 kHz (dashed), (ii)
$\Gamma_{\mathrm{R}}/2\pi$ = 4.3 kHz (continuous), and (iii) $\Gamma_{\mathrm{R}}/2\pi$ = 5.2 kHz (dotted), with the rest of the parameters
the same as in the continuous fit in (a). (c) Predictions from our model for different values of the unequal feeding parameter $\beta$: (i)
$\beta$ = 0 (dashed), (ii) $\beta$ = 0.1 (continuous), (iii) $\beta$ = 0.5 (dotted), with the rest of the parameters the same as in the
continuous fit in (a). (d) Predictions from our model for different values of the initial transmission: (i) $T_0$ = 0.4 (dashed), (ii) $T_0$
= 0.46 (continuous), and (iii) $T_0$ = 0.6 (dotted), with the rest of the parameters the same as in the continuous fit in (a). The
continuous curve is the same in all the figures.}
\end{figure}

Figure \ref{fig5}(a) shows the measured evolution of the group delay (through a He* cell of length 2.5 cm) versus coupling beam intensity
for a beam diameter of 1.5 cm (dots), reproduced from Ref.\,\cite{goldfarb08}. The results have been obtained with Gaussian probe pulses of
duration of 70 $\mu$s with a peak power of 35 $\mu$W and with the coupling and probe beam frequencies at the center of the Doppler profile
($\Delta = \delta_{\mathrm{R}} = 0$).

In the slow-light experiments performed with He*, reported in Ref.\,\cite{goldfarb08}, the measured delays for non-zero coupling intensities
were slightly overestimated as the reference used was the probe pulse in the absence of a coupling beam.

Theoretically, the group delay is calculated from the derivative of the real part of the susceptibility with respect to frequency at
two-photon resonance, for the experimentally used parameters. The continuous curve in Fig.\,\ref{fig5}(a) is the best fit obtained from our
model with $\Gamma_{\mathrm{R}}/2\pi$ = 4.3 kHz, $\Gamma_{\mathrm{t}}/2 \pi$ = 0.41 kHz, $\eta = 10^{-4}$ and $\beta$ = 0.1, for the same
probe power of 35 $\mu$W and the same initial transmission $T_0$ = 0.46 as in the experiment.

The group velocity derived \cite{goldfarb08} from the susceptibility of Ref.\,\cite{lvovsky06} leads to a group delay at the line center
($\Delta = \delta_{\mathrm{R}} = 0$) given by:
\begin{equation}
\tau_{\mathrm{g}}=-\ln(T_0)\frac{(2W_{\mathrm{D}}+ \Gamma)\Omega_{\mathrm{C}}^2}{\left[2\Gamma_{\mathrm{R}}(2W_{\mathrm{D}}+ \Gamma)
+\Omega_{\mathrm{C}}^2\right]^2}\;. \label{lvovsky_delay}
\end{equation}
The maximum value of the group delay is reached for $\Omega_{\mathrm{C}}^2 = 2\Gamma_{\mathrm{R}}(2W_{\mathrm{D}}+\Gamma)$ and its value is
$-\ln(T_0)/8\Gamma_{\mathrm{R}}$. Again, there is a better agreement of the experimental results with our present theory than what was
obtained in Ref.\,\cite{goldfarb08} using Eq.\,(\ref{lvovsky_delay}), with the value of $\Gamma_{\mathrm{R}}$ deduced from the EIT-width
data.

Figure \ref{fig5}(a) also shows the comparison of the group delay profiles from our theory for different values of $\eta = 10^{-3}$ (dashed
curve) and 0 (dotted curve) around the best-fit value corresponding to $\eta = 10^{-4}$ (continuous curve), for the 1.5 cm-diameter beam.
The magnified part shown in the inset justifies the choice of the VCC parameter value of $\eta = 10^{-4}$ for the measured data for He*,
which is distinct from that of $\eta = 0$. As mentioned earlier, the measured data were over-estimated and a good fit running slightly below
the data points would require a sensitive adjustment of $\eta > 0$ (as shown).

Figure \ref{fig5}(b) shows the comparison of the group delay profiles from the theory for different values of $\Gamma_{\mathrm{R}}$ around
the best-fit value, for the 1.5 cm-diameter beam. $\Gamma_{\mathrm{R}}$ indeed has a great impact on the delay which is clear because lower
the value of $\Gamma_{\mathrm{R}}$ (for example, the dashed curve for $\Gamma_{\mathrm{R}}/2\pi$ = 3.2 kHz), more is the coherence, hence
higher are the delays achieved in the system.

Figure \ref{fig5}(c) shows the same as above for different values of $\beta$ around the best-fit value, for a range of $\beta$ from 0 to
0.5. We see that with an increase of $\beta$, the delays at different coupling intensities decrease. On comparison with the plots of peak
transmission for different values of $\beta$ in Fig.\,\ref{fig4}(c), one deduces that the effect of $\beta$ on these two characteristics are
opposite. We can understand this as follows. If an increase of $\beta$ increases the peak transmission (as seen and explained above), then
from the relationship of resonant peak transmission and delay, it becomes clear that an increased transmission would lead to a decrease in
the group delay. This is also manifested in Eqs.\,(\ref{lvovsky_trans}) and (\ref{lvovsky_delay}). The inset at a magnified scale shows the
comparison of the plots for $\beta = 0$ and $\beta = 0.1$ with reference to the measured data points -- it is clear that $\beta = 0.1$
provides a distinctly better fit than $\beta = 0$.

Figure \ref{fig5}(d) shows the comparison of the group delay profiles from the theory for different values of $T_0$ around the best-fit
value, for the 1.5 cm-diameter beam. The group delay increases with a decrease in $T_0$ and the effect is also supported by
Eq.\,(\ref{lvovsky_delay}). A decrease in the initial transmission in the absence of the coupling field signifies a proportionate increase
in the number of participating atoms. $T_0$, however, does not affect the EIT width. Thus the delay-bandwidth product, which is a figure of
merit for a delay/storage medium, turns out to be proportional to the number density of the medium.

In comparison to the width and the peak transmission plots shown in Figs.\,\ref{fig3} and \ref{fig4} respectively, we note that
$\Gamma_{\mathrm{R}}$, $\eta$, $\beta$ and $T_0$ have a greater impact on the delay as seen from Fig.\,\ref{fig5}, especially in determining
the peak of the delay curve.

\section{Conclusions}

We have presented a realistic analysis of EIT and slow light in a hot atomic vapor, taking into account all the relevant decoherence
processes, for arbitrary strengths of the probe and the coupling fields, and without assuming any special initial condition. We have
considered the influx of fresh atoms in the lower levels and the outflux from all the levels at a transit rate in the gas. Unlike the theory
for EIT in a Doppler-broadened medium in Ref.\,\cite{javan02} in which the role of collisions is completely neglected, our analysis includes
phase-interrupting collisions of active atoms as well as velocity-changing collisions, modeled effectively in the strong collision limit. We
have demonstrated the role of VCCs in redistributing the atomic population over all velocity classes and hence a velocity-independent
optical pumping. The initial condition of $\rho_{bb}$ = 1, i.e., a system perfectly prepared by optical pumping at $t$ = 0 in
Ref.\,\cite{lvovsky06}, is redundant here in the presence of rapid VCC coverage.

The steady-state solutions for the atomic density matrix elements are presented in the strong-collision approximation with a model for the
VCCs under rapid VCC coverage. A value of the VCC parameter $\eta >0$ indicates a loss of coherence by VCCs. As observed from the results,
all the EIT characteristics have a sensitive dependence on $\eta$. In our example of the He* system \cite{goldfarb08}, it was found that a
small value of $\eta = 10^{-4}$ gives the best fits for all the measured characteristics for different beam sizes.

For He*, the motion of the atoms through the beam is assumed to be diffusive, because of the large number of collisions they undergo.
Further, because of diffusion and favourable collisions outside the interaction region, we have allowed a slightly greater fraction of atoms
to enter the beam prepared in state $\vert b \rangle$ than those in state $\vert c \rangle$ using a parameter $\beta$. The best fit value of
$\beta > 0$ supports this fact. For all systems in which other decoherence effects are small so that atoms can diffuse out of the
interaction region and return before decohering, the unequal feedback fraction would model such a positive contribution to the coherence,
similar in effect to a decrease in the number density or an increase in the initial transmission $T_0$. A constant value of $\beta$ is found
to fit different beam sizes, given the precision of the reported measurements.

For non-zero detunings $\Delta$ of the coupling field from the center of the Doppler-broadened transition frequency, the transmitted
intensity profiles become asymmetric about the two-photon resonance (Raman detuning $\delta_{\mathrm{R}}$ = 0). This Fano-like feature is a
signature of the two-photon process of EIT, and emerges naturally from our model. The EIT width $\Gamma_{\mathrm{EIT}}$, simulated from our
model, shows a linear dependence on the coupling beam intensity, i.e., a quadratic dependence on the coupling beam Rabi frequency
$\Omega_{\mathrm{C}}$, as observed in experiments. The evolutions of the peak transmission and the group delay with the coupling beam
intensity predicted from our analysis faithfully reproduce the experimentally observed behaviors. In the evolution of all these features of
interest, an increase in the Raman decoherence rate $\Gamma_{\mathrm{R}}$ seems in general to have a similar effect as an increase in the
VCC parameter $\eta$. But whatever the beam diameter, $\eta$ should remain constant, while $\Gamma_{\mathrm{t}}$ and hence
$\Gamma_{\mathrm{R}}$ should decrease when the beam diameter increases (keeping $\Gamma_{\mathrm{R}}-\Gamma_{\mathrm{t}}$ constant).

EIT has recently found applications in white-light cavities \cite{wu08} for use in gravitational wave detection. For applications of slowing
of light for use in quantum-information processing, and in particular, in quantum memory, the medium needs to be optically dense. Our model
is applicable to all such systems of hot atomic vapor, which are attractive candidates for practical applications requiring large-bandwidth
controllable delays, for example, for signal processing at the appropriate wavelength. The system can be generalized to model tripod-like
systems \cite{tripod} in hot vapor.

\begin{acknowledgments}

This work is supported by an Indo-French Networking Project funded by the Department of Science and Technology, Government of India, and the
French Ministry of Foreign Affairs. The work was initiated during a stay of RG in France, supported also by ``C'Nano Ile-de-France'' and
``Triangle de la Physique''. The authors wish to thank E. Arimondo, H. Gilles, D. Kumar, M. Leduc, P. J. Nacher and G. Tastevin for useful
discussions. The work of JG is supported by the Council of Scientific and Industrial Research, India. The School of Physical Sciences,
Jawaharlal Nehru University, is supported by the University Grants Commission, India, under a Departmental Research Support scheme.

\end{acknowledgments}


\begin{thebibliography}{99}
\bibitem{approx} In both Refs.\,\cite{javan02} and \cite{lvovsky06}, in order to avoid the complexity in the integration with a
Maxwell-Boltzmann velocity distribution, the usual Gaussian distribution has been approximated with a Lorentzian function, leading to small deviations.
\bibitem{Rabi} Our Rabi frequency is the real angular frequency of the Rabi oscillations, and is thus double of the one defined in Refs.\,\cite{javan02} and \cite{lvovsky06}.
\bibitem{comment} Other than the anomaly related to the actual beam size inside the cell as mentioned in the text, note that the three sets of data for the EIT width and the two sets of data for peak transmission in Ref.\,\cite{goldfarb08} were generated under different experimental conditions, notably the initial transmission $T_0$, signifying differences in the number density of the participating atoms in the He* discharge cell. Likewise, the transit time of the atoms in the beam, influenced by collisions, and the collisional dephasing $\Gamma_{\mathrm{coh}}^p$, would be different for these different measurements. The best fits from our theory are consistent with the logical adjustments of both $T_0$ and $\Gamma_{\mathrm{R}}$ for these different sets of data.
\bibitem{f} Note that in the calculation of the control field Rabi frequency $\Omega_{\mathrm{C}}$ in Ref.\,\cite{goldfarb08}, the oscillator strength for the transition has been erroneously taken as the total $f = 0.54$ instead of $f/2$ (in case of degenerate Zeeman sublevels, when a circularly-polarized light probes the transition along any one arm of the $\Lambda$-system, the oscillator strength is half of the total $f$.). It has resulted in a scaling of $\Omega_{\mathrm{C}}$, which has been corrected here. This has no bearing on the conclusions drawn in Ref.\,\cite{goldfarb08}.
\end{thebibliography}
\end{document}